\documentclass[%
reprint,
 superscriptaddress,
 amsmath,amssymb,
 aps,
prb,
]{revtex4-2}

\usepackage{multirow}
\usepackage{array}

\usepackage[colorlinks=true,bookmarks=false,citecolor=blue,urlcolor=blue]{hyperref}

\usepackage{xcolor}
\usepackage{float}
\usepackage{url}
\usepackage{graphicx}
\usepackage{bm}
\usepackage{ulem}

\begin{document}

\preprint{APS/123-QED}

\title{Analytical model of metasurfaces comprising meta-atoms with anisotropic polarizabilities}


\author{Izzatjon Allayarov}
\affiliation{Hannover Centre for Optical Technologies, Leibniz University Hannover, 30167 Hannover, Germany}
\affiliation{Institute of Transport and Automation Technology, Leibniz University Hannover, 30823 Garbsen, Germany}
\affiliation{Cluster of Excellence PhoenixD, Leibniz University Hannover, 30167 Hannover, Germany}

\author{Vladimir R. Tuz}
\affiliation{School of Radiophysics, Biomedical Electronics and Computer Systems, V. N. Karazin Kharkiv National University, 4 Svobody Square, Kharkiv 61022, Ukraine}

\author{Antonio Cal{\`a} Lesina}
\affiliation{Hannover Centre for Optical Technologies, Leibniz University Hannover, 30167 Hannover, Germany}
\affiliation{Institute of Transport and Automation Technology, Leibniz University Hannover, 30823 Garbsen, Germany}
\affiliation{Cluster of Excellence PhoenixD, Leibniz University Hannover, 30167 Hannover, Germany}

\author{Andrey B. Evlyukhin}%
\affiliation{Cluster of Excellence PhoenixD, Leibniz University Hannover, 30167 Hannover, Germany}
\affiliation{Institute of Quantum Optics, Leibniz University Hannover, 30167 Hannover, Germany}

\date{\today}

\begin{abstract}
A general analytical approach to the study of electromagnetic resonances of metasurfaces consisting of meta-atoms with anisotropic electric and magnetic dipole polarizabilities and irradiated with obliquely incident light is developed  in the direct dipole-moment representation. The presented approach allows us to clearly trace and explain the features of the electromagnetic coupling between electric and magnetic dipole moments in the metasurface and to identify its role in the formation of optical resonances. For these purposes, the dependence of the dipole lattice sums on the angle of illumination is also presented. 
 Expressions for the transmittance  and reflectivity corresponding to the specular reflectance  are presented with explicit inclusion of the dipole moments of particles in the array, which allows using these expressions for multipole analysis of purely numerical results concerning the optical response of metasurfaces. The developed analytical method is tested to characterize the spectral resonances of dielectric metasurfaces composed of rectangular silicon nanoprisms. In addition, we discuss the relationship and similarity between the results of coupled dipole and coupled dipole-quadrupole methods. Our analytical representation of electromagnetic dipole coupling is an insightful and fast method for the characterization of collective resonances in metasurfaces under illumination at oblique incidence. It could be especially useful for designing planar nanophotonic devices consisting of arbitrary-shaped building blocks and operating under special irradiation conditions.

\end{abstract}

\maketitle

\section{\label{sec:into} Introduction}

Periodic arrays of nanoparticles also known as metasurfaces have many applications, for example in structural color generation~\cite{yang2020all,wang2023structural}, flat optics~\cite{chen2020flat,shastri2023nonlocal}, nonlinear~\cite{abir2022second,vabishchevich2023nonlinear} and quantum photonics~\cite{solntsev2021metasurfaces,santiago2022resonant} and holography~\cite{ni2013metasurface,huang2018metasurface}. Designing and further optimization of metasurfaces for such applications require a lot of computational resources and repetitive simulations over a set of parameters. In this context, analytical and semi-analytical approaches can help to predict and better understand their spectral features with less simulation resources. In many cases, the above applications are based on exploiting the electric and magnetic dipole resonances of plasmonic and dielectric/semiconductor metasurfaces as well as the coupling between them occurring under different irradiation conditions. Therefore, dipole approximation approaches are important for analysis \cite{mulholland1994light,merchiers2007light}. In the literature, there are several formulations of the coupled-dipole method (CDM), which are obtained for an array of spherical particles under normal~\cite{garcia2007colloquium,evlyukhin2010optical} and oblique incidence~\cite{abujetas2018generalized,Abujetas2020coupled}. They are successfully used to study surface lattice resonances~\cite{babicheva2017resonant,allayarov2024dynamic}, the Brewster effect \cite{abujetas2018generalized}, symmetry protected and accidental bound states in the continuum (BICs)~\cite{Murai2020bound,abujetas2021near,abujetas2022tailoring}.

In this paper, in addition to the existing formulations, we consider an important  special case by extending the CDM to the case of metasurfaces composed of meta-atoms with anisotropic polarizabilities and illuminated at an arbitrary incidence angle. In that case single particle electric and magnetic dipole polarizabilities are considered to be not scalars anymore, but are  2-rank tensors with different nonzero diagonal elements. Accordingly, the optical response of the metasurface depends not only on the polarization, but also on the choice of plane and angle of incidence of external light waves. Moreover, when light falls obliquely on the metasurface, electromagnetic coupling occurs between electric and magnetic dipoles, regardless of the symmetric properties of the elementary cells. And this in turn can lead to the emergence of collective resonances with interference features.

The paper is organized as follows: in Sec.~\ref{sec:dmodel}, we introduce a coupled system of equations for electric and magnetic dipole moments and lattice sums. After discussing the dependence of the lattice sums on the incidence angle in Sec.~\ref{sec:lattice}, we proceed to Sec.~\ref{sec:tesec} and Sec.~\ref{sec:tmsec}, where we solve the coupled-dipole equations for TE and TM polarization, respectively, and derive reflection and transmission coefficients in terms of dipole moments. Similarities between the CDM and the coupled dipole-quadrupole model (CDQM) in terms of eigenmode conditions (i.e., BICs) under normal incidence are discussed in Sec.~\ref{sec:sec_condBIC}. A demonstrative comparison of our analytical model with full numerical simulation results is presented and discussed in Sec.~\ref{sec:res}. Finally, conclusions are drawn in Sec.~\ref{sec:concl}.

\begin{figure}[b]
\centering
\includegraphics[width=1\linewidth]{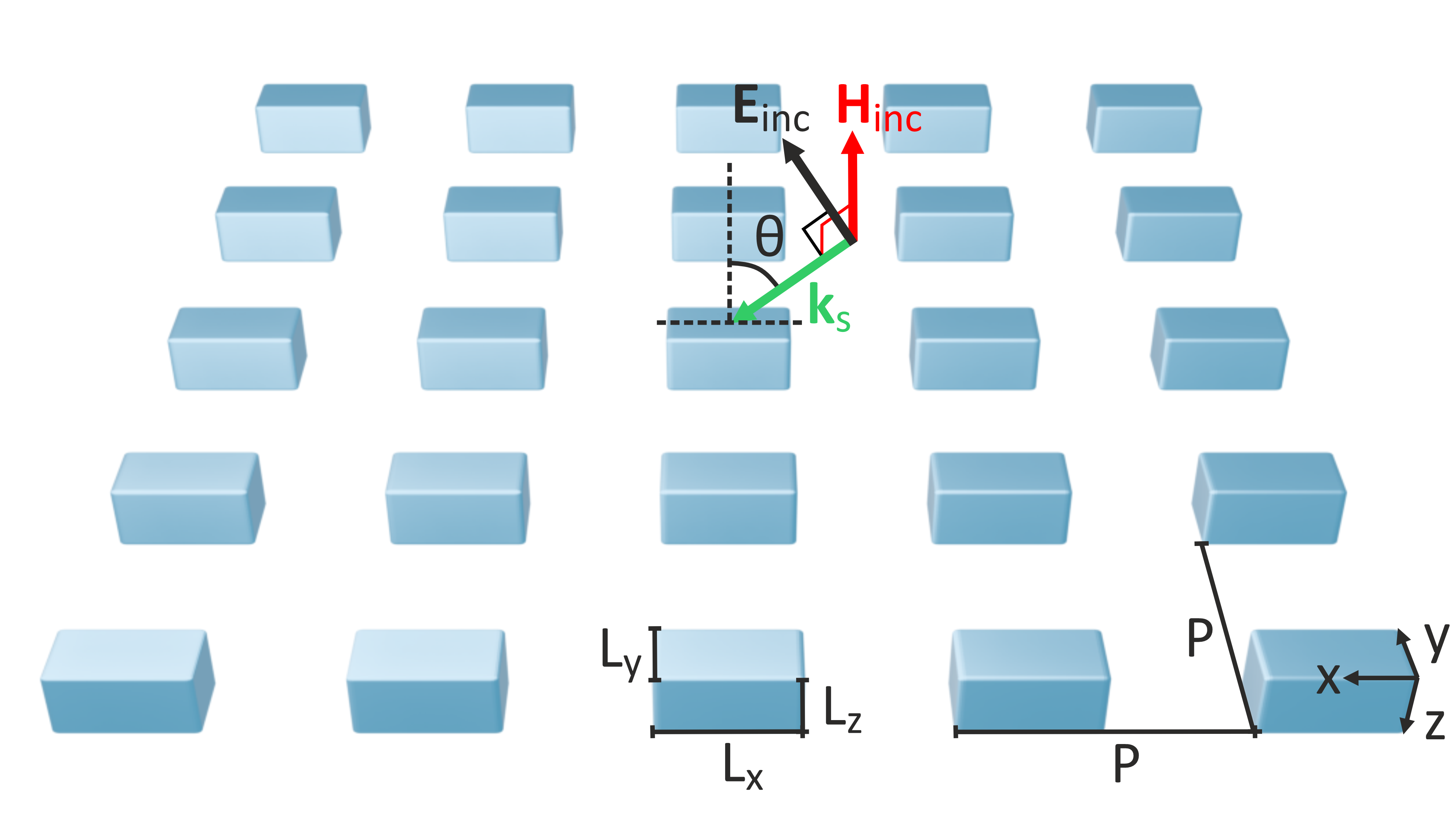}
\caption{Schematics of a metasurface made of rectangular nanoprisms and respective irradiation condition. The plane of incidence is the $xz$-plane corresponding to the in-plane angle  $\varphi=0$. The incident angle is denoted by $\theta$. The origin of the coordinate system should be in the center of mass of a particle.}
\label{fig:str}
\end{figure}

\section{Coupled-dipole model\label{sec:dmodel}}

Let us consider an infinite 2D periodic array of identical nanoparticles (rectangular nanoprism) in a homogeneous environment with relative permittivity $\varepsilon_{\rm s}$ (Fig.~\ref{fig:str}). For the convenience of our derivation, we assume that the centers of mass of the nanoparticles are in the $xy$-plane at $z=0$. The array is excited with a linearly polarized plane wave at oblique incidence. Hence, based on Ref.~\cite{evlyukhin2010optical} and using Bloch's theorem as in Ref.~\cite{garcia2007colloquium}, the coupled-dipole equations for the electric (ED) ${\bf p}$ and magnetic (MD) ${\bf m}$ dipole moments of the nanoparticle at the lattice node with the number ${l}=0$ and located at the origin of the Cartesian coordinate system (coordinate vector ${\bf r}_0=0$) can be written as
\begin{align}
    &{\bf p}=\varepsilon_0\varepsilon_{\rm s}\hat\alpha^{\rm p}{\bf E}_0+\hat\alpha^{\rm p}\hat S{\bf p}+\hat\alpha^{\rm p}\frac{{\rm i}}{v}[{\bf g}\times{\bf m}],\label{pmsys1}\\
    &{\bf m}=\hat\alpha^{\rm m}{\bf H}_0+\hat\alpha^{\rm m}\hat S{\bf m}+\hat\alpha^{\rm m}\frac{v}{{\rm i}}[{\bf g}\times{\bf p}], \label{pmsys2}
\end{align}
where $\varepsilon_0$ is the vacuum dielectric constant,
${\bf E}_0$ and ${\bf H}_0$ are the electric and magnetic fields of the incident wave at the origin of the Cartesian system, respectively,  $v=c/n_{\rm s}$ is the speed of light in the surrounding medium with refractive index $n_{\rm s}$ ($\varepsilon_{\rm s}=n_{\rm s}^2$), $\hat\alpha^{\rm p}$ and  $\hat\alpha^{\rm m}$  are the electric and magnetic dipole polarizability tensors, respectively, and `i' is the imaginary unit. The lattice sums in the above equations are presented as 
 \begin{equation}
    \hat S=k_{\rm s}^2\sum_{l\neq0}\hat G(0,{\bf r}_l){\rm e}^{{\rm i}{\bf k_{\parallel}}{\bf r}_l},\quad
    {\bf g}=k_{\rm s}\sum_{l\neq0} {\bf \Tilde{g}}(0,{\bf r}_l){\rm e}^{{\rm i}{\bf k_{\parallel}}{\bf r}_l},\label{Sg}
\end{equation}
 where ${\bf k}_{\parallel}$ is the component of the incoming wave
vector parallel to the array, ${\bf r}_{l}\ne 0$ is the radius vector of the $l$-th lattice node, where the ED  ${\bf p}_l={\bf p}\exp({\rm i}{\bf k}_{\parallel}{\bf r}_l)$ and MD ${\bf m}_l={\bf m}\exp({\rm i}{\bf k}_{\parallel}{\bf r}_l)$ of a particle with number $l$ are located.
 Here, the tensor $\hat{G}$ and vector ${\bf \Tilde{g}}$ are \cite{evlyukhin2010optical}
\begin{align}
  \hat{G}(0, {\bf r}_l)= &\frac{{\rm e}^{{\rm i} k_{\rm s} { r}_l}}{4 \pi { r}_l} \left [ \left(1 + \frac{{\rm i}}{k_{\rm s}{ r}_l} -\frac{1}{k_{\rm s}^2 { r}_l^2}  \right) \hat{U}\right.\nonumber\\
  &\left.+ \left(-1 - \frac{3{\rm i}}{k_{\rm s} { r}_l} + \frac{3}{k_{\rm s}^2 { r}_l^2} \right) \mathbf{n} \mathbf{n} \right ],\\
  {\bf \Tilde{g}}(0,{\bf r}_l)=&\frac{{\rm e}^{{\rm i}k_{\rm s}{ r}_l}}{4\pi { r}_l}\left(\frac{1}{{ r}_l^2}-\frac{{\rm i}k_{\rm s}}{{ r}_l}\right){\bf r}_l,
\end{align}
where $r_l=|{\bf r}_l|$, ${\bf n}=-{\bf r}_l/{ r}_l$, $k_{\rm s}=k_0n_{\rm s}$ is the wave-number in the surrounding medium. For a nanobar supporting a resonant dipole response, the tensors of electric $\hat\alpha^{\rm p}$ and magnetic $\hat\alpha^{\rm m}$ polarizabilities can be written as
\begin{equation}\label{alpha}
\hat\alpha^{\rm p}=\left( {\begin{matrix}
   {\alpha^{\rm p}_{x}} & 0 & {0} \cr
    0& {\alpha^{\rm p}_{y}} & {0} \cr
   {0} & {0} & {\alpha^{\rm p}_{z}} \cr
\end{matrix}
} \right), \qquad
\hat\alpha^{\rm m}=\left( {\begin{matrix}
   {\alpha^{\rm m}_{x}} & 0 & {0} \cr
    0& {\alpha^{\rm m}_{y}} & {0} \cr
   {0} & {0} & {\alpha^{\rm m}_{z}} \cr
\end{matrix}
} \right),
\end{equation}
in the coordinate system shown in Fig.~\ref{fig:str}. Here, we consider that the single particle polarizability's components  do not depend on the irradiation directions (the local response)~\cite{bobylev2020nonlocal}. {Note that  the  tensors $\hat\alpha^{\rm p}$ and $\hat\alpha^{\rm m}$ can be determined by material and geometrical anisotropy of meta-atoms. In general, they are calculated numerically (for details, see Appendix~\ref{sec:appA}). However, for a homogeneous spherical particle, they can be analytically determined within the Mie theory~\cite{babicheva2024mie}, for an ellipsoid in the quasi-static regime can also be expressed analytically~\cite{evlyukhin2005surface}.}

In the coordinate system shown in Fig.~\ref{fig:str}, the electric, magnetic and wave vector of the incident wave are ${\bf E}_{\rm inc}={\bf E}_0\,{\rm exp}({{\rm i}{\bf k_{\rm s}}{\bf r}})$, ${\bf H}_{\rm inc}={\bf H}_0\,{\rm exp}({{\rm i}{\bf k_{\rm s}}{\bf r}})$ and ${\bf k}_{\rm s}=k_{\rm s}(\sin\theta,0,\cos\theta)$, respectively. We define the components of ${\bf E}_0$ and ${\bf H}_0$ for TE polarization incidence as
\begin{align}\label{te1}
    &{\bf E}_0=E_0(0,1,0), \\
    &{\bf H}_0=H_0(-\cos\theta,0,\sin\theta),
\end{align}
and for TM polarization as
\begin{align}\label{tm1}
    &{\bf E}_0=E_0(\cos\theta,0,-\sin\theta),\\
    &{\bf H}_0=H_0(0,1,0).
\end{align}
Before solving Eqs.~(\ref{pmsys1}) and (\ref{pmsys2}) for both polarizations, let us look at the components of the vector ${\bf g}$ and tensor $\hat{S}$. For the periodic system shown in Fig.~\ref{fig:str} and chosen irradiation conditions with ${\bf k}_\parallel=k_x=k_s\sin\theta$, one can show that due to symmetry only $g_x$, and diagonal elements of $\hat{S}$, i.e., $S_x\equiv S_{xx}$, $S_y\equiv S_{yy}$ and $S_z\equiv S_{zz}$ are nonzero~\cite{evlyukhin2010optical}:
\begin{align}
     &g_x=k_{\rm s}\sum_{l\neq0} x_l F_l \left(\frac{1}{r_l^2}-\frac{{\rm i}k_{\rm s}}{r_l}\right) \ne 0\quad {\rm if}\quad {\sin\theta}\ne 0,\label{gx}\\
     &S_{\beta}\!=\!k_{\rm s}^2\!\sum\limits_{l\ne0}\!F_l\!\left(\!1\!+\!\frac{{\rm i}}{k_{\rm s}r_l}\!-\!\frac{1}{k_{\rm s}^2r_l^2}\!-\!\frac{\beta_l^2}{r_l^2}\!-\!\frac{3{\rm i}\beta_l^2}{k_{\rm s}r_l^3}\!+\!\frac{3\beta_l^2}{k_{\rm s}^2r_l^4}\!\right),\label{S}
\end{align}
where  $F_l = {\rm exp}[{\rm i}k_{\rm s}(r_l+x_l\sin\theta)]/(4\pi r_l)$, $r_l=|{\bf r}_l|=\sqrt{x_l^2+y_l^2}$, and $\beta$ stands for $x$, $y$ and $z$. Remind that we consider all dipoles located in the $xy$-plane so that  $z_l=0$ for any $l$.  Furthermore, note that the above sums are calculated over all lattice nodes excluding the node with $l=0$ at the origin of the coordinate system, where the  ED $\bf p$ and MD $\bf m$ of the particle with number $l=0$  are located.

\subsection{Angular dependence of the lattice sums \label{sec:lattice}}

\begin{figure}[h]
\centering
\includegraphics[width=1\linewidth]{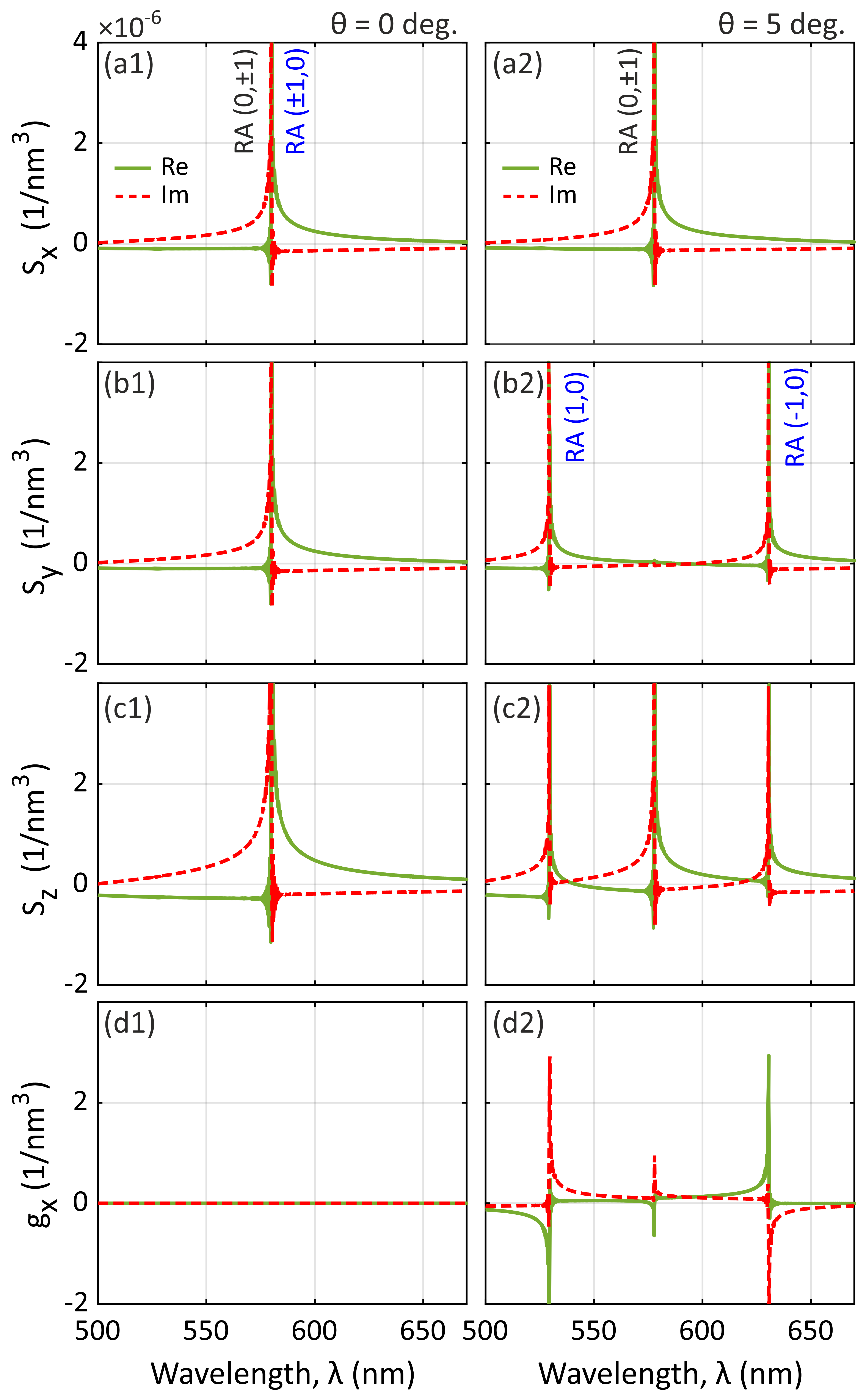}
\caption{Real and imaginary parts of the lattice sums $S_x$, $S_y$, $S_z$ and $g_x$ as a function of wavelength for (a1-d1) $\theta=0^{\circ}$ (normal incidence) and (a2-d2) $\theta=5^{\circ}$. The square lattice has the period of $P=400$~nm. The surrounding refractive index is $n_{\rm s}=1.45$. The legends shown in (a1,a2) are valid for all the panels below. Spectral positions of the Rayleigh anomaly (RA) orders RA($n_x$,$n_y$) are defined as $\lambda=Pn_{\rm s}(\sqrt{n_x^2+n_y^2\cos^2\theta}-n_x\sin\theta)/(n_x^2+n_y^2)$,  where $(n_x,n_y) \in Z$ that is, they are integers~\cite{ndao2018plasmonless}. The values of the wavelength $\lambda$
 are indicated for vacuum.}
\label{fig:sum}
\end{figure}

To demonstrate what changes occur to the spectra of the lattice sums in Eqs.~(\ref{gx}) and (\ref{S}) for excitation at oblique incidence, we write them in the following form:
\begin{equation}\label{gg}
    g_x=\frac{{\rm i}k_{\rm s}}{2\pi}\sum_{x_l>0} x_l\sin(x_lk_{\rm s}\sin\theta)\frac{e^{{\rm i}k_{\rm s}r_l}}{r_l}\left(\frac{1}{r_l^2}-\frac{{\rm i}k_{\rm s}}{r_l}\right),
\end{equation}
\begin{equation}
    S_z=\sum_{l\ne0}\cos(x_lk_{\rm s}\sin\theta)\frac{e^{{\rm i}k_{\rm s}r_l}}{4\pi r_l}\left(k_{\rm s}^2+\frac{{\rm i }k_{\rm s}^2}{k_{\rm s}r_l}-\frac{1}{r_l^2}\right),
\end{equation}
\begin{equation}\label{Sx}
    S_x=S_z+\sum_{l\ne0}x_l^2\cos(x_lk_{\rm s}\sin\theta)\frac{e^{{\rm i}k_{\rm s}r_l}}{4\pi r_l}\left(\frac{3}{r_l^4}-\frac{k_{\rm s}^2}{r_l^2}-\frac{{3\rm ik_{\rm s}}}{r_l^3}\right),
\end{equation}
\begin{equation}\label{Sy}
    S_y=S_z+\sum_{l\ne0}y_l^2\cos(x_lk_{\rm s}\sin\theta)\frac{e^{{\rm i}k_{\rm s}r_l}}{4\pi r_l}\left(\frac{3}{r_l^4}-\frac{k_{\rm s}^2}{r_l^2}-\frac{{3\rm ik_{\rm s}}}{r_l^3}\right).
\end{equation}
The results of the calculations of these sums for a certain set of parameters and two angles of incidence are presented in Fig.~\ref{fig:sum}. It can be seen that in the selected spectral range, at normal incidence ($\theta=0^{\rm o}$), the sum $g_x$ is equal to zero, while the sums $S_x$, $S_y$ and $S_z$ have a divergence only at one point, $\lambda=Pn_{\rm s}=\lambda^{\rm RA}$, which corresponds to the  Rayleigh anomaly (RA). For oblique incidence ($\theta=5^{\rm o}$), the resonant behavior of the lattice sums changes significantly, as shown in Figs.~\ref{fig:sum}(a2)-\ref{fig:sum}(d2): $S_z$ and $g_x$ have three resonances, $S_y$ has two resonances, and $S_x$ has only one resonance, similar to the case of normal illumination.

Understanding the spectral behavior of the resonances  in Fig.~\ref{fig:sum}  follows from the corresponding  far-field (FF) terms of $S_z$
\begin{equation}\label{SzFF_1}
    S_z^{\rm FF}|_{x_l=0}=\frac{k_{\rm s}^2}{4\pi}\sum_{l\ne0,x_l=0}\frac{e^{{\rm i}k_{\rm s}|y_l|}}{|y_l|},
\end{equation}
and
\begin{equation}\label{SzFF_2}
    S_z^{\rm FF}|_{y_l=0}=\frac{k_{\rm s}^2}{4\pi}\sum_{l\ne0,y_l=0}\cos(x_lk_{\rm s}\sin\theta)\frac{e^{{\rm i}k_{\rm s}|x_l|}}{|x_l|}.
\end{equation}
The term (\ref{SzFF_1}), calculated only for $x_l=0$, does not depend on the angle of incidence $\theta$ and therefore its divergence determines the middle resonance of $S_z$ in Fig.~\ref{fig:sum}(c2), which occurs at the wavelength $\lambda^{\rm RA}$, as in the case of normal incidence shown in Fig.~\ref{fig:sum}(c1). 
The term (\ref{SzFF_2}), calculated only for $y_l=0$, depends on the angle of incidence $\theta$ due to the factor $\cos(x_lk_s\sin\theta)$, as for the chain of particles from Ref.~\cite{markel2005divergence}. Therefore, its divergence will be at the spectral points determined by $(1\pm\sin\theta) Pk_s=2\pi L$, $L$
 being an integer. In the selected section of the spectrum in Fig.~\ref{fig:sum}, $L=1$, and we obtain two divergence points $\lambda=Pn_s(1\pm\sin\theta)$ that correspond to two lateral resonances in Fig~\ref{fig:sum}(c2).

Using (\ref{SzFF_1}) and (\ref{SzFF_2}) one can obtain that the divergent far-field parts of (\ref{Sx}) and (\ref{Sy}) are
\begin{eqnarray}\label{SxFF}
    S_x^{\rm FF}|_{y_l=0}&=&0,\nonumber\\
    S_x^{\rm FF}|_{x_l=0}&=&\frac{k_{\rm s}^2}{4\pi}\sum_{l\ne0,x_l=0}\frac{e^{{\rm i}k_{\rm s}|y_l|}}{|y_l|},
\end{eqnarray}
and
\begin{eqnarray}\label{SyFF}
    S_y^{\rm FF}|_{y_l=0}&=&\frac{k_{\rm s}^2}{4\pi}\sum_{l\ne0,y_l=0}\cos(x_lk_{\rm s}\sin\theta)\frac{e^{{\rm i}k_{\rm s}|x_l|}}{|x_l|},\nonumber\\
    S_y^{\rm FF}|_{x_l=0}&=&0\:.
\end{eqnarray}
Thus, comparing (\ref{SxFF}) and (\ref{SxFF})  with (\ref{SzFF_1}) and (\ref{SzFF_2}) we obtain that $S_X$ C has one point of divergence, while $S_y$ has two points of divergence in full accordance with Fig.~\ref{fig:sum}(a2) and Fig.~\ref{fig:sum}(b2), respectively. A similar analysis of (\ref{gg}) leads to the result shown in Fig ~\ref{fig:sum}(d2).

Here, we would like to note that in real space, the convergence of the numerical calculations of the elements of $\hat{S}$ is slow~\cite{Abujetas2020coupled}. However, by using the approach from the Appendix of Ref.~\cite{evlyukhin2010optical}, one can derive the following analytical expressions, at least for the imaginary parts of $\hat{S}$ in the diffractionless regime:
\begin{align}
    &{\rm Im}(S_x)=\frac{k_{\rm s}\cos\theta}{2S_{\rm L}}-\frac{k_{\rm s}^3}{6\pi},\label{imSx}\\
    &{\rm Im}(S_y)=\frac{k_{\rm s}}{2S_{\rm L}\cos\theta}-\frac{k_{\rm s}^3}{6\pi},\label{imSy}\\
    &{\rm Im}(S_z)=\frac{k_{\rm s}\sin\theta\tan\theta}{2S_{\rm L}}-\frac{k_{\rm s}^3}{6\pi},\label{imSz}\\
    &{\rm Im}({\rm i}g_x)=-\frac{k_{\rm s}}{2S_{\rm L}}\tan\theta,\label{regx}
\end{align}
where $S_{\rm L}$ is the area of the lattice unit cell.

\subsection{TE polarization \label{sec:tesec}}

The coupled-dipole Eqs.~(\ref{pmsys1}) and (\ref{pmsys2}) for TE polarization reduce to the following:
\begin{align}
    &{ p}_y=\varepsilon_0\varepsilon_{\rm s}\alpha^{\rm p}_{y}{ E}_0+\alpha^{\rm p}_{y} S_y{ p}_y-\alpha^{\rm p}_{y}\frac{{\rm i}}{v}g_xm_z,\label{py}\\
    &{ m}_x=-\alpha^{\rm m}_{x}{ H}_0\cos\theta+\alpha^{\rm m}_{x} S_x{ m}_x,\label{mx}\\
    &{ m}_z=\alpha^{\rm m}_{z}{ H}_0\sin\theta+\alpha^{\rm m}_{z} S_z{ m}_z+\alpha^{\rm m}_{z}\frac{v}{{\rm i}}{ g}_x{p}_y.\label{mz}
\end{align}
As evident from Eqs.~\eqref{py} and \eqref{mz}, under oblique incidence conditions, there is coupling (energy exchange) between the electric and magnetic dipole subsystems, which disappears under normal incidence \cite{evlyukhin2010optical}. The solution of Eq.~\eqref{mx} is
\begin{equation}\label{mxet}
    m_x=-\frac{H_0\cos\theta}{1/\alpha^{\rm m}_x-S_x},
\end{equation}
while the solutions of the system of Eqs.~\eqref{py} and \eqref{mz} for $p_y$ and $m_z$ can be written as
\begin{align}
  &p_y=\frac{(1/\alpha^{\rm m}_z-S_z)-{\rm i}g_x\sin\theta}{({1}/{\alpha^{\rm p}_y}-S_y) ({1}/{\alpha^{\rm m}_z}-S_z)+g_x^2}  \varepsilon_0\varepsilon_{\rm s}E_0, \label{pet}\\
  &m_z=\frac{(1/\alpha^{\rm p}_y-S_y)\sin\theta-{\rm i}g_x}{({1}/{\alpha^{\rm p}_y}-S_y) ({1}/{\alpha^{\rm m}_z}-S_z)+g_x^2} H_0. \label{met}
\end{align}
The zeros of the determinant of the system for a certain incidence angle $\theta_0$, defines the conditions for eigenmodes including bound states in the continuum (BIC) with purely real eigenvalues (see Sec.~\ref{sec:sec_condBIC}):
\begin{equation}\label{bicTE}
    \left[\frac{1}{\alpha^{\rm p}_y}-S_y(\theta_0)\right] \left[\frac{1}{\alpha^{\rm m}_z}-S_z(\theta_0)\right]+g_x^2(\theta_0)=0.
\end{equation}

To calculate the reflection $r^{\rm TE}$ and transmission $t^{\rm TE}$ coefficients for the TE polarized incidence, it is more convenient to use the electric field since it has only one nonzero component. The total electric field ${\bf E}$ at a point $\bf r$ outside the metasurface is a sum of the incident electric field ${\bf E}_{\rm inc}$ and the electric fields generated by all EDs and MDs: 
\begin{align}
    &{\bf E}_{\rm ED}({\bf r})=\dfrac{k_0^2}{\varepsilon_0}\sum_{l=0}^\infty\hat{G}({\bf r}, {\bf r}_l){\rm e}^{{\rm i}{\bf k_{\parallel}}{\bf r}_l}{\bf p},\label{EED}\\
    &{\bf E}_{\rm MD}({\bf r})=\dfrac{{\rm i}}{\varepsilon_0 c}\sum_{l=0}^\infty[{\bf \Tilde{g}}({\bf r},{\bf r}_l){\rm e}^{{\rm i}{\bf k_{\parallel}}{\bf r}_l}\times {\bf m}],\label{EMD}
\end{align}
 respectively. Thus, the total electric field $E_y({\bf r})$ is
\begin{eqnarray}\label{Etot}
    E_y({\bf r})&=&E_0{\rm e}^{{\rm i}k_{\rm s}(x\sin\theta +z\cos\theta)}\nonumber\\
    &&+\frac{k^2_0}{\varepsilon_0}\biggl(\sum_{l=0}^\infty G_{yy}({\bf r}, {\bf r}_l){\rm e}^{{\rm i}x_lk_{\rm s}\sin\theta}\biggr)p_y\nonumber\\
    &&+\frac{{\rm i}k_0}{\varepsilon_0 c}\biggl(\sum_{l=0}^\infty \Tilde{g}_z({\bf r},{\bf r}_l){\rm e}^{{\rm i}x_lk_{\rm s}\sin\theta}\biggr)m_x\nonumber\\
    &&-\frac{{\rm i}k_0}{\varepsilon_0 c}\biggl(\sum_{l=0}^\infty \Tilde{g}_x({\bf r},{\bf r}_l){\rm e}^{{\rm i}x_lk_{\rm s}\sin\theta}\biggr)m_z.
\end{eqnarray}                                                                                        
The real space sums in Eq.~\eqref{Etot} can be written in the reciprocal space by using the approach from the Appendix of Ref.~\cite{evlyukhin2010optical}:
\begin{align}
   &\sum_{l=0}^\infty \! G_{yy}({\bf r}, {\bf r}_l){\rm e}^{{\rm i}x_lk_{\rm s}\sin\theta} \!=\!\sum_{\bf L}^\infty\frac{{\rm i}(k_{\rm s}^2\!-\!L_y^2){\rm e}^{{\rm i}\xi|z|-{\rm i}\sigma}}{2S_{\rm L} k^2_{\rm s}\xi},\label{sum1}\\
   &\sum_{l=0}^\infty \Tilde{g}_z({\bf r},{\bf r}_l){\rm e}^{{\rm i}x_lk_{\rm s}\sin\theta}=\sum_{\bf L}^\infty \frac{\pm {\rm e}^{{\rm i}\xi|z|-{\rm i}\sigma}}{2S_{\rm L}},\label{sum2}\\
   &\sum_{l=0}^\infty \Tilde{g}_x({\bf r},{\bf r}_l){\rm e}^{{\rm i}x_lk_{\rm s}\sin\theta}\!=\!\sum_{\bf L}^\infty\frac{(L_x\!-\!k_{\rm s}\sin\theta) {\rm e}^{{\rm i}\xi|z|-{\rm i}\sigma}}{2S_{\rm L}\xi},\label{sum3}
\end{align}
where $\bf L$ denotes the reciprocal-lattice vectors, in Eq.~\eqref{sum2}, `$+$' corresponds to $z<0$ and `$-$' to $z>0$, $\sigma \!=\! (L_x-k_{\rm s}\sin\theta)x+L_y y$, $\xi \!=\! [{k_{\rm s}^2\!-\!(L_x\!-\!k_{\rm s}\sin\theta)^2\!-\!L_y^2}]^{1/2}$, and $x$, $y$, $z$ are spatial coordinates.

Further, we focus on the case of ${\bf L}=0$, i.e., zero diffraction order (the specular reflectance). Thus, by substituting Eqs.~\eqref{sum1}-\eqref{sum3} into Eq.~\eqref{Etot}, one obtains the following reflected ($z<0$, `R') and transmitted ($z>0$, `T') scattered fields in the far-field region, respectively
\begin{equation}
 E_y^{{\rm R}}\!=\!\frac{{\rm i}k_{\rm s}{\rm e}^{{\rm i}k_{\rm s}(x\sin\theta-z\cos\theta)}}{2S_{\rm L}\varepsilon_0\varepsilon_{\rm s}\cos\theta}\!\biggl(p_y\!+\!\frac{\cos\theta}{v}m_x\!+\!\frac{\sin\theta}{v}m_z\!\biggr), \label{Eref}
\end{equation}
\begin{align}
 &E_y^{{\rm T}}=E_0{\rm e}^{{\rm i}k_{\rm s}(x\sin\theta+z\cos\theta)} \nonumber \\
 &\!+\! \frac{{\rm i}k_{\rm s}{\rm e}^{{\rm i}k_{\rm s}(x\sin\theta+z\cos\theta)}}{2S_{\rm L}\varepsilon_0\varepsilon_{\rm s}\cos\theta}\biggl(p_y\!-\!\frac{\cos\theta}{v}m_x\!+\!\frac{\sin\theta}{v}m_z\biggr). \label{Etra}
\end{align}

 The corresponding field reflection coefficient is $r^{\rm TE}\!=\!E_y^{\rm R}(z\!=\!0)/E_0{\rm e}^{-{\rm i}k_{\rm s}x\sin\theta}$, i.e.,
\begin{equation}\label{re}
  r^{\rm TE} \!=\! \frac{{\rm i}k_{\rm s}}{2S_{\rm L}\varepsilon_0\varepsilon_{\rm s} E_0\cos\theta }\!\left(p_y\!+\!\frac{\cos\theta}{v}m_x\!+\!\frac{\sin\theta}{v}m_z\!\right),  
\end{equation}
and the field transmission coefficient is $t^{\rm TE}\!=\!E_y^{\rm T}(z\!=\!0)/E_0{\rm e}^{-{\rm i}k_{\rm s}x\sin\theta}$, i.e.,
\begin{equation}\label{te}
  t^{\rm TE} \!=\! 1 \!+\!\frac{{\rm i}k_{\rm s}}{2S_{\rm L}\varepsilon_0\varepsilon_{\rm s} E_0\cos\theta }\!\biggl(\!p_y\!-\!\frac{\cos\theta}{v}m_x\!+\!\frac{\sin\theta}{v}m_z\!\!\biggr).
\end{equation}
It is important to note that in Eqs.~\eqref{re} and \eqref{te}, $p_y$, $m_x$, and $m_z$ correspond only to the dipole moments located at the origin of the coordinate system.

By using solutions (\ref{mxet})-(\ref{met}) and also introducing the following quantities,
\begin{align}
    &1/\tilde{\alpha}^{\rm m}_x=({1/\alpha^{\rm m}_x-S_x})/\cos\theta, \label{at_TE1} \\
    &1/\tilde{\alpha}^{\rm p}_y=({1/\alpha^{\rm p}_y-S_y})\sin\theta, \label{at_TE2} \\
    &1/\tilde{\alpha}^{\rm m}_z=({1/\alpha^{\rm m}_z-S_z})/\sin\theta, \label{at_TE3} \\
    &{S}^{\prime}_{\rm L}=S_{\rm L}\cos\theta,
\end{align}
one can represent $r^{\rm TE}$ and $t^{\rm TE}$ as $r^{\rm TE} =r^{\rm TE}_{\rm coup}-r^{\rm TE}_{\rm nocoup}$ and $t^{\rm TE}=1 + r^{\rm TE}_{\rm coup}+r^{\rm TE}_{\rm nocoup}$, where
\begin{align}
 &r^{\rm TE}_{\rm coup} \!=\! \frac{{\rm i}k_{\rm s}}{2{S}^{\prime}_{\rm L}}\!\Bigg[\frac{(1/\tilde{\alpha}^{\rm p}_y \!-\! {\rm i}g_x)\!+\! (1/\tilde{\alpha}^{\rm m}_z \!-\! {\rm i}g_x)}{1/(\tilde{\alpha}^{\rm p}_y\tilde{\alpha}^{\rm m}_z) + g^2_x}\Bigg]\sin\theta, \label{re1_coupl} \\
 &r^{\rm TE}_{\rm nocoup} =\frac{{\rm i}k_{\rm s}}{2{S}^{\prime}_{\rm L}}\tilde{\alpha}^{\rm m}_x\cos\theta. \label{re1_nocoupl}
\end{align}
{are the coupled and uncoupled parts of the reflection coefficient, respectively.}

\subsection{TM polarization \label{sec:tmsec}}

For the TM polarization, Eqs.~(\ref{pmsys1}) and (\ref{pmsys2}) can be written as
\begin{align}
    &{ p}_x=\varepsilon_0\varepsilon_{\rm s}\alpha^{\rm p}_x{ E}_0\cos\theta+\alpha^{\rm p}_x S_x{ p}_x,\label{px}\\
    &{ m}_y=\alpha^{\rm m}_y{ H}_0 +\alpha^{\rm m}_y S_y{ m}_y-\frac{v}{{\rm i}}\alpha^{\rm m}_y{ g}_x{p}_z,\label{my}\\
    &{ p}_z=-\varepsilon_0\varepsilon_{\rm s}\alpha^{\rm p}_z{ E}_0\sin\theta+\alpha^{\rm p}_z S_z{ p}_z+\frac{{\rm i}}{v}\alpha^{\rm p}_z{ g}_x{m}_y.\label{pz}
\end{align}

The solutions of the above equations for $p_x$, $m_y$, and $p_z$ are
\begin{align}
  &p_x=\frac{\varepsilon_0\varepsilon_{\rm s}E_0\cos\theta}{1/\alpha^{\rm p}_x-S_x}, \label{ptm}\\
  &m_y=\frac{(1/\alpha^{\rm p}_z-S_z)-{\rm i}g_x\sin\theta}{(1/\alpha^{\rm m}_y-S_y) (1/\alpha^{\rm p}_z-S_z)+g_x^2} H_0, \label{mtm}\\
  &p_z=\frac{-(1/\alpha^{\rm m}_y-S_y)\sin\theta+{\rm i}g_x}{(1/\alpha^{\rm m}_y-S_y) (1/\alpha^{\rm p}_z-S_z)+g_x^2} \varepsilon_0\varepsilon_{\rm s}E_0. \label{ptmz}
\end{align}
The eigenmode conditions for the TM polarization case can be found from
\begin{equation}\label{bicTM}
    \left[\frac{1}{\alpha^{\rm m}_y}-S_y(\theta_0)\right] \left[\frac{1}{\alpha^{\rm p}_z}-S_z(\theta_0)\right]+g_x^2(\theta_0)=0.
\end{equation}

In the case of TM polarization, it is convenient to calculate the field reflection $r^{\rm TM}$ and transmission $t^{\rm TM}$ coefficients via the magnetic field. The total magnetic  field ${\bf H}$ at a point $\bf r$ outside the metasurface is a sum of the incident magnetic field ${\bf H}_{\rm inc}$ and the electric fields generated by all EDs and MDs:
\begin{align}
    &{\bf H}_{\rm ED}({\bf r})={k_{\rm s}^2}\sum_{l=0}^\infty\hat{G}({\bf r}, {\bf r}_l){\rm e}^{{\rm i}{\bf k_{||}}{\bf r}_l}{\bf m},\label{HED}\\
    &{\bf H}_{\rm MD}({\bf r})={-{\rm i}v}\sum_{l=0}^\infty[{\bf \Tilde{g}}({\bf r},{\bf r}_l){\rm e}^{{\rm i}{\bf k_{||}}{\bf r}_l}\times {\bf p}],\label{HMD}
\end{align}
 respectively. Thus, the total magnetic field $H_y({\bf r})$ is
\begin{eqnarray}\label{Htot}
    H_y({\bf r})&=&H_0{\rm e}^{{\rm i}k_{\rm s}(x\sin\theta +z\cos\theta)}\nonumber\\
    &&+{k_{\rm s}^2}\biggl(\sum_{l=0}^\infty G_{yy}({\bf r}, {\bf r}_l){\rm e}^{{\rm i}x_lk_{\rm s}\sin\theta}\biggr)m_y\nonumber\\
    &&-{{\rm i}v}\biggl(\sum_{l=0}^\infty \Tilde{g}_z({\bf r},{\bf r}_l){\rm e}^{{\rm i}x_lk_{\rm s}\sin\theta}\biggr)p_x\nonumber\\
    &&+{{\rm i}v}\biggl(\sum_{l=0}^\infty \Tilde{g}_x({\bf r},{\bf r}_l){\rm e}^{{\rm i}x_lk_{\rm s}\sin\theta}\biggr)p_z.
\end{eqnarray}
Using an analogy with the case of the TE polarization, we obtain the following reflected and transmitted magnetic fields:
\begin{equation}
 H_y^{\rm R}({\bf r})=\frac{{\rm i}k_{\rm s}{\rm e}^{{\rm i}k_{\rm s}(x\sin\theta\!-\!z\cos\theta)}}{2S_{\rm L}\varepsilon_0\varepsilon_{\rm s}\cos\theta}\left(m_y\!-\!vp_x\cos\theta\!-\!vp_z\sin\theta\right).
\end{equation}
\begin{align}
 &H_y^{\rm T}({\bf r})=H_0{\rm e}^{{\rm i}k_{\rm s}(x\sin\theta +z\cos\theta)}\nonumber\\
 &+\frac{{\rm i}k_{\rm s}{\rm e}^{{\rm i}k_{\rm s}(x\sin\theta+z\cos\theta)}}{2S_{\rm L}\varepsilon_0\varepsilon_{\rm s}\cos\theta}\left(m_y\!+\!vp_x\cos\theta\!-\!vp_z\sin\theta\right).
\end{align}
Hence, the reflection and transmission coefficients (into the zero diffraction order) are
\begin{align}
  &r^{\rm TM} = \frac{{\rm i}k_{\rm s}}{2S_{\rm L}H_0\cos\theta }\left(m_y-vp_x\cos\theta-vp_z\sin\theta\right),\label{rh}\\ 
  &t^{\rm TM}\!=\!1 \!+ \!\frac{{\rm i}k_{\rm s}}{2S_{\rm L} H_0\cos\theta }\left(m_y\!+\!vp_x\cos\theta\!-\!vp_z\sin\theta\right). \label{th}
\end{align}
Introducing the following quantities
\begin{align}
    &1/\tilde{\alpha}^{\rm p}_x=({1/\alpha^{\rm p}_x-S_x})/\cos\theta, \label{at_TM1} \\
    &1/\tilde{\alpha}^{\rm p}_z=({1/\alpha^{\rm p}_z-S_z})/\sin\theta, \label{at_TM2} \\
    &1/\tilde{\alpha}^{\rm m}_y=({1/\alpha^{\rm m}_y-S_y})\sin\theta, \label{at_TM3}
\end{align}
and using Eqs. (\ref{ptm})-(\ref{ptmz}) the coefficients  $r^{\rm TM}$ and $t^{\rm TM}$ can be rewritten as $r^{\rm TM} =r^{\rm TM}_{\rm coup}-r^{\rm TM}_{\rm nocoup}$ and $t^{\rm TM}=1 + r^{\rm TM}_{\rm coup}+r^{\rm TM}_{\rm nocoup}$, where
\begin{align}
  &r^{\rm TM}_{\rm coup} \!=\! \frac{{\rm i}k_{\rm s}}{2{S}^{\prime}_{\rm L}}\!\Bigg[\frac{(1/\tilde{\alpha}^{\rm p}_z \!-\! {\rm i}g_x) + (1/\tilde{\alpha}^{\rm m}_y \!-\! {\rm i}g_x)}{1/(\tilde{\alpha}^{\rm p}_z\tilde{\alpha}^{\rm m}_y) + g^2_x} \!\Bigg]\!\sin\theta, \label{rh1_coup} \\
  &r^{\rm TM}_{\rm nocoup} = \frac{{\rm i}k_{\rm s}}{2{S}^{\prime}_{\rm L}}\tilde{\alpha}^{\rm p}_x\cos\theta. \label{rh1_nocoup}
\end{align}

For the both TE and TM polarizations, the intensity reflectance $R$, transmittance $T$ and absorbance $A$ are defined as $R=|r|^2$, $T=|t|^2$ and $A=1-R-T$, respectively. In the case of diffraction, the coefficient A also includes radiation into non-zero diffraction orders.

\subsection{\label{sec:sec_condBIC} Conditions for BICs}

\begin{table*}[t]
\caption{\label{tab:table1} This table lists similarities between the coupled-dipole model (CDM) and coupled-dipole-quadrupole model under normal incidence conditions (CDQM-normal). For CDQM-normal, the excitation conditions and notations are as in Ref.~\cite{allayarov2024anapole}.}
\begin{ruledtabular}
\begin{tabular}{lllll}
 Model & CDM & CDQM-normal & CDM
& CDQM-normal\\ \hline
 Subsystem & TE & Even & TM & Odd \\
 Coupling parameter & $g_x$ & $-S_{\rm Qm}$ & $g_x$ & $S_{\rm Mp}$\\
 Coupled moments & $p_y$ and $m_z$ & $m_y$ and $Q_{xz}$ & $m_y$ and $p_z$  & $p_x$ and $M_{yz}$\\
 aBIC condition & $1/\tilde{\alpha}^{\rm p}_y = 1/{\tilde{\alpha}^{\rm m}_z} = {\rm i}g_x$ & $1/{\tilde{\alpha}_{\rm m}}=1/{\tilde{\alpha}_{\rm Q}}= -S_{\rm Qm}$ & $1/{\tilde{\alpha}^{\rm m}_y} = 1/{\tilde{\alpha}^{\rm p}_z} = {\rm i}g_x$ & $1/{\tilde{\alpha}_{\rm p}}=1/{\tilde{\alpha}_{\rm M}}= S_{\rm Mp}$\\
 Resonant part of reflection & $r^{\rm TE}_{\rm coup}$ & $r_{\rm even}$ & $r^{\rm TM}_{\rm coup}$ & $r_{\rm odd}$\\
\end{tabular}
\end{ruledtabular}
\end{table*}

The eigenmode conditions in Eqs.~\eqref{bicTE} and \eqref{bicTM} are general and can define spectral positions of both symmetry protected (at $\theta=0$) and accidental (at $\theta \neq 0$) BICs. Indeed, at normal incidence $g_x=0$ (there is no coupling between in-plane and out-of-plane dipole components) and if we consider, for example, Eq.~\eqref{bicTE}, it can be written as $(1/{\alpha^{\rm p}_y}-S_y)(1/\alpha^{\rm m}_z-S_z)=0$. The expression in the first pair of brackets cannot have zeros (in the diffractionless region), since for the non-absorbing case  one can analytically show that ${\rm Im}(1/{\alpha^{\rm p}_y}-S_y)=-{k_{\rm s}}/({2S_{\rm L}})\neq0$~\cite{allayarov2024anapole}. While zeros of its real part, i.e., ${\rm Re}(1/{\alpha^{\rm p}_y}-S_y)=0$ define spectral positions of in-plane ED surface lattice resonance (SLR)~\cite{garcia2007colloquium,allayarov2024dynamic}.

On the other hand, the zeros of the expression in the second pair of brackets can be defined via the real part's solutions since for the non-absorbing case  its imaginary part ${\rm Im}(1/{\alpha^{\rm m}_z}-S_z)=0$ in the diffractionless region. Thus, for the TE polarization, the condition for the spectral position of the symmetry-protected BIC (sBIC) is ${\rm Re}(1/\alpha^{\rm m}_{z}-S_z)=0$ \cite{babicheva2021multipole}. A similar analysis can be done for the TM polarization case and obtain the condition for sBIC, ${\rm Re}(1/\alpha^{\rm p}_z-S_z)=0$.

At $\theta \neq 0$, the coupling parameter $g_x\neq0$, and there is a coupling between in-plane and out-of-plane dipole components. Hence, one can expect that at certain conditions these radiating dipoles cancel each other and the state becomes a non-radiating eigenstate, which is known as accidental BIC (aBIC)~\cite{azzam2021photonic,abujetas2022tailoring}. Indeed, by using Eqs.~(\ref{at_TE2}) and (\ref{at_TE3}), and Eqs.~(\ref{at_TM2}) and (\ref{at_TM3}), one can rewrite the conditions in Eq.~\eqref{bicTE} and Eq.~\eqref{bicTM} as the following:
\begin{align}
    &1/(\tilde{\alpha}^{\rm p}_y {\tilde{\alpha}^{\rm m}_z}) - ({\rm i}g_x)^2 = 0,\quad \text{TE polarization}, \label{bicTE1} \\
    &1/({\tilde{\alpha}^{\rm p}_z} {\tilde{\alpha}^{\rm m}_y}) - ({\rm i}g_x)^2 = 0,\quad \text{TM polarization}. \label{bicTM1}
\end{align}
The formal solutions of the above equations are
\begin{align}
    &1/\tilde{\alpha}^{\rm p}_y = 1/{\tilde{\alpha}^{\rm m}_z} = \pm {\rm i}g_x,\quad \text{TE polarization}, \label{eigcondTE}\\
    &1/{\tilde{\alpha}^{\rm p}_z} = 1/{\tilde{\alpha}^{\rm m}_y} = \pm {\rm i}g_x,\quad \text{TM polarization}. \label{eigcondTM}
\end{align}
By using Eqs.~(\ref{imSx})-(\ref{regx}) and the derivations from the Supplemental Material of Ref.~\cite{allayarov2024anapole}, one can show that the imaginary parts of the above solutions are the same in the diffractionless spectral region, i.e.,
\begin{align}
     &{\rm Im}(1/\tilde{\alpha}^{\rm p}_y) = {\rm Im}(1/{\tilde{\alpha}^{\rm m}_z}) ={\rm Im}({\rm i}g_x) = -\frac{k_{\rm s}}{2S_{\rm L}}\tan\theta, \label{eigcondTE_im} \\
     &{\rm Im}(1/{\tilde{\alpha}^{\rm p}_z}) = {\rm Im}(1/{\tilde{\alpha}^{\rm m}_y}) = {\rm Im}({\rm i}g_x) = -\frac{k_{\rm s}}{2S_{\rm L}}\tan\theta. \label{eigcondTM_im}
\end{align}
Therefore, only `+' in (\ref{eigcondTE}) and  (\ref{eigcondTM}) corresponds to the physically correct solutions. As a result,  the conditions for aBIC are
\begin{align}
    &{\rm Re}(1/\tilde{\alpha}^{\rm p}_y) = {\rm Re}(1/{\tilde{\alpha}^{\rm m}_z}) = {\rm Re}({\rm i}g_x),\quad \text{TE polarization}, \label{eigcondTE_re} \\
    &{\rm Re}(1/{\tilde{\alpha}^{\rm p}_z}) = {\rm Re}(1/{\tilde{\alpha}^{\rm m}_y}) = {\rm Re}({\rm i}g_x),\quad \text{TM polrization}. \label{eigcondTM_re}
\end{align}

Here, we would like to note the similarity between the above-derived coupled-dipole model results for oblique incidence and dipole-quadrupole model for normal incidence. As it has been shown in Ref.~\cite{allayarov2024anapole}, within the coupled dipole-quadrupole model, for certain parameters, normally excited metasurfaces can exhibit aBIC as a result of the destructive interference of resonant either odd (e.g., electric dipole and magnetic quadrupole) or even (e.g., magnetic dipole and electric quadrupole) parity multipoles. As one can see from Table~\ref{tab:table1}, the conditions for aBIC are very similar (other elements as well). This is because in an array, the contributions of out-of-plane dipole components to the radiation from the array can be considered analogous to the contributions of certain components of the quadrupole tensors. Furthermore, as it is shown in Refs.~\cite{allayarov2024multiresonant,allayarov2024anapole}, the system with high-order multipoles can be treated within the effective dipole approach.

\section{\label{sec:res} Testing and Discussion}

\begin{figure}[b]
\centering
\includegraphics[width=1\linewidth]{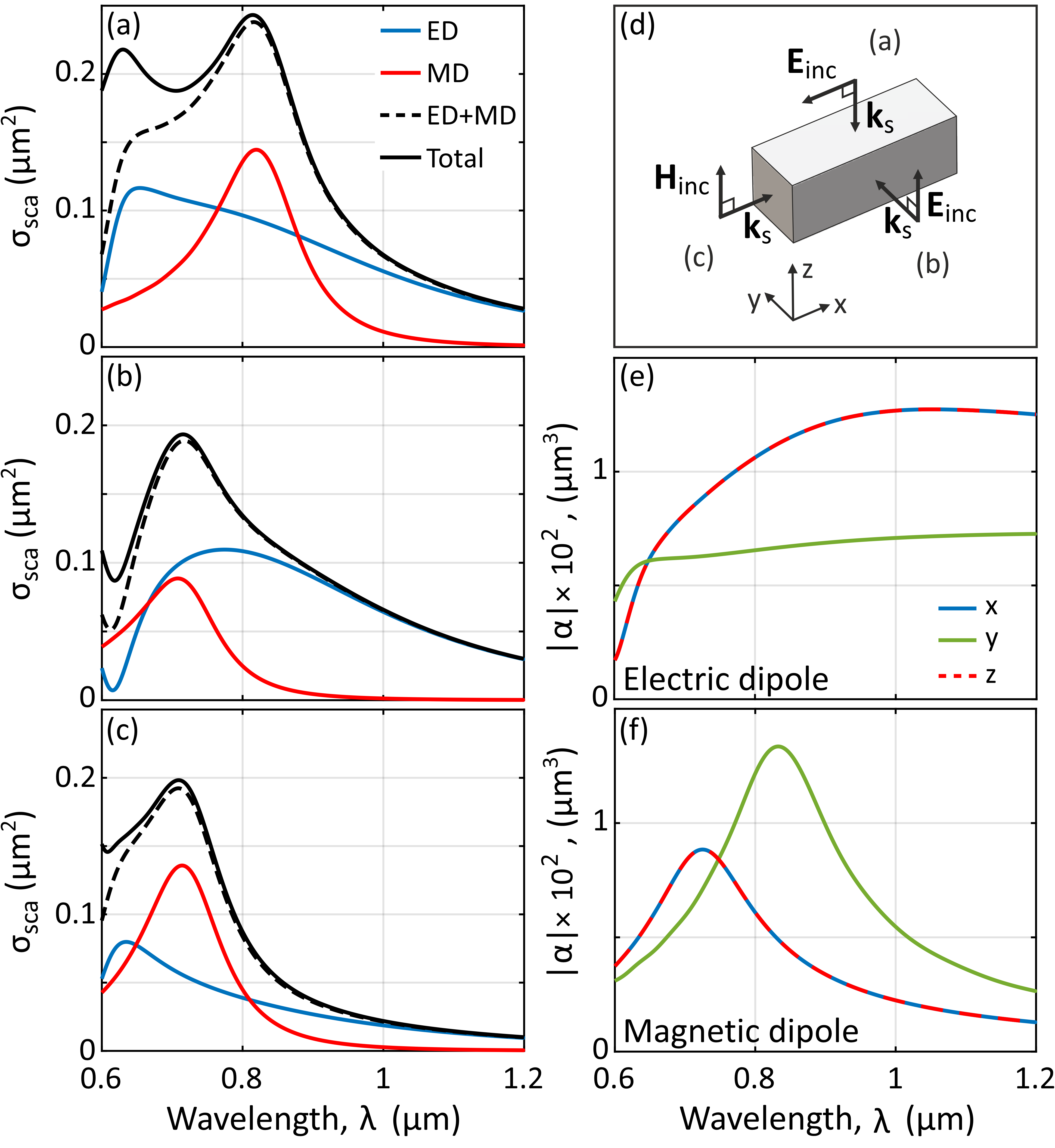}
\caption{(a-c) Total scattering cross-section of a single silicon nanoprism and its dipole contributions as a function of the wavelength for the excitation conditions shown in the panel (d). Modulus of the averaged components of the (e) electric and (f) magnetic dipole polarizabilities of the single nanoprism. The values of the wavelength $\lambda$
 are indicated for vacuum.}
\label{fig:single}
\end{figure}

To demonstrate the applicability of our analytical model, let us consider the metasurface presented in Fig.~\ref{fig:str}. We assume that the rectangular nanoprisms are made of silicon~\cite{green2008self} and immersed in a homogeneous environment with a refractive index of $n_{\rm s}=1.45$, which approximately corresponds to glass. The nanoprisms' width, height, and thickness are $L_x=200$~nm, $L_y=115$~nm, and $L_z=200$~nm, respectively. For these parameters, a single nanoprism  exhibits a dipole response. Indeed, as one can see from Figs.~\ref{fig:single}(a)-(c), the single particle has electric and magnetic dipole resonances in the considered spectral range for three different excitation conditions shown in Fig.~\ref{fig:single}(d). Furthermore, based on similar simulations (for details see Appendix~\ref{sec:appA}), one can numerically calculate the averaged components of anisotropic dipole polarizability tensors $\hat\alpha$ [see Eq.~(\ref{alpha})], which are shown in Figs.~\ref{fig:single}(e) and \ref{fig:single}(f). We note that $\alpha_x=\alpha_z$ since $L_x=L_z$.

\begin{figure}[b]
\centering
\includegraphics[width=0.9\linewidth]{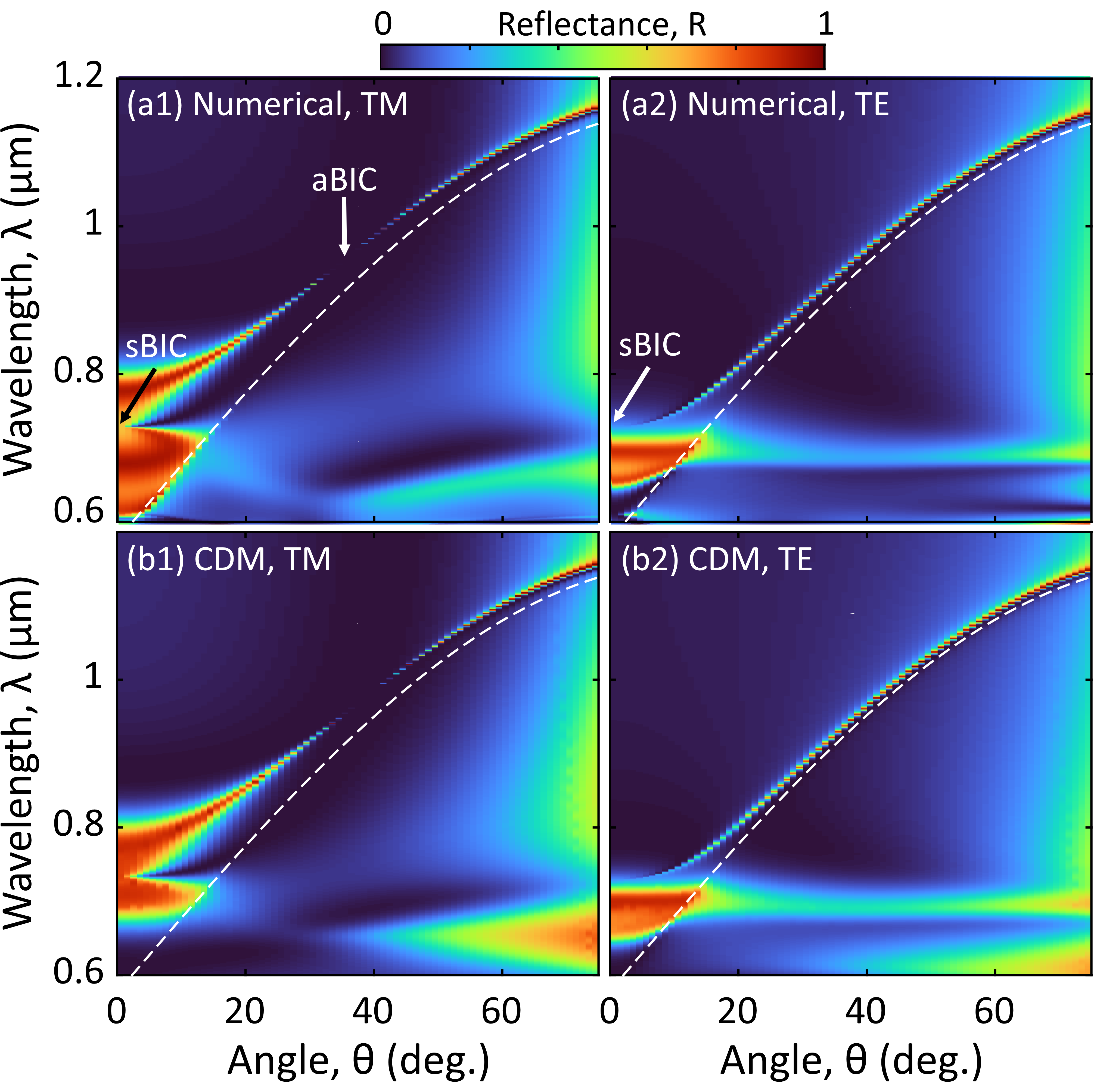}
\caption{(a1-a2) Full numerical and (b1-b2) coupled-dipole model (CDM) specular  reflectance (the zero diffraction order) of the metasurface for (a1,b1) TM and (a2,b2) TE polarizations as a function of the incidence polar angle $\theta$. The sketch of the metasurface is shown in Fig.~\ref{fig:str} and the parameters of particles are as in Fig.~\ref{fig:single}. The plane of incidence is $xz$-plane. The period of the metasurface is $P=400$~nm in both directions. The white dashed lines indicate RA(-1,0) position, defined as $\lambda=Pn_{\rm s}(1+\sin\theta)$.  sBIC: symmetry protected BIC; aBIC: accidental BIC. The values of the wavelength $\lambda$ are indicated for
vacuum.}
\label{fig:array_xz}
\end{figure}

\begin{figure}[t]
\centering
\includegraphics[width=0.9\linewidth]{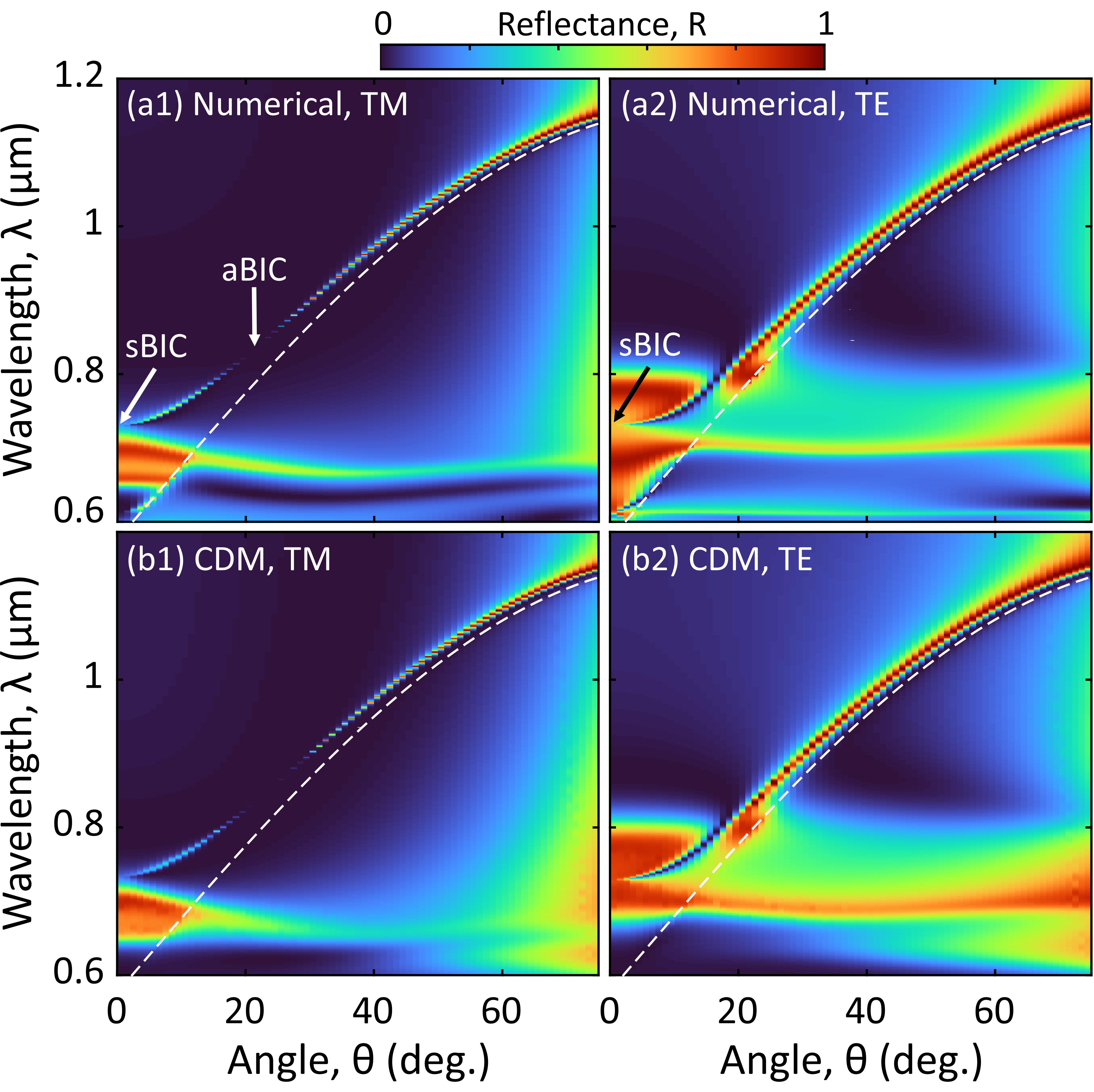}
\caption{The same reflectance calculations as in Fig.~\ref{fig:array_xz}, but for the $yz$-plane of incidence.}
\label{fig:array_yz}
\end{figure} 

In addition to the angle-independent polarizabilities, one can easily calculate the lattice sums in Eqs.~(\ref{gx}) and (\ref{S}) for the interested incidence polar angles. These two quantities, i.e., single particle polarizabilities and lattice sums are enough to identify spectral features of the metasurface irradiated at an angle. Figures~\ref{fig:array_xz}(a1) and \ref{fig:array_xz}(a2) depict the numerically simulated reflectance of the metasurface as a function of the polar angle for the TM and TE polarized excitation in the $xz$-plane of incidence. The system exhibits sBIC for both polarizations at ($\theta=0^{\circ}$, $\lambda=730$~nm), and aBIC at the position ($\theta=35^{\circ}$, $\lambda=950$~nm) for the TM polarization. The results based on the analytical coupled-dipole model are presented in Figs.~\ref{fig:array_xz}(b1) and \ref{fig:array_xz}(b2). One can see that the results of the CDM have  very good agreement with the full-wave numerical calculations performed in the ANSYS Lumerical's RCWA solver. The difference at lower wavelengths is due to quadrupoles, which are not included in the CDM. This can be seen from Figs.~\ref{fig:single}(a)-(c), in which the difference between solid and dashed black lines corresponds to quadrupole contributions.

If we change the plane of incidence from $xz$ to $yz$, the lattice sums do not change. However, in-plane components of the polarizability tensors interchange, i.e., $\alpha_ x^{yz}=\alpha_y^{xz}$, $\alpha_y^{yz}=\alpha_x^{xz}$. Full numerical and CDM results for the $yz$-plane of incidence are shown in Fig.~\ref{fig:array_yz}. Again, the model shows very good agreement with numerical simulation results, although due to the anisotropy of the single particle dipole polarizabilities, the reflectance is changed. One can note that the positions of sBICs remain unchanged, since $\alpha_z^{yz}=\alpha_z^{xz}$. On the other hand, in this case, the aBIC occurs at the position ($\theta=20^{\circ}$, $\lambda=835$~nm). This is due to the fact that conditions for aBIC [see Eqs.~(\ref{eigcondTE_re}) and (\ref{eigcondTM_re})] include an in-plane component of polarizabilities, the value of which depends on the plane of incidence.

\section{Conclusion \label{sec:concl}}

In conclusion, we have derived a general representation of the coupled-dipole approach that can be applied to metasurfaces consisting of nanoparticles with an anisotropic polarizability. Furthermore, in our analytical model, specular reflection and transmission coefficients are expressed in terms of dipole moments, which is more convenient for analysis of pure numerically obtained results since dipole moments can be also  directly numerically calculated (e.g., with the multipole decomposition approach). A comparison of our coupled-dipole model results with full numerical simulations showed very good agreement between them. Additionally, we demonstrated an explicit dependence of lattice sums on the angle of illumination and direction of interparticle coupling in the array. We believe that the results obtained in this paper are useful for a deeper understanding of collective resonances and coupling effects in metasurfaces and for their applications in nanophotonics.

\section*{\label{sec:acknl} Acknowledgments}
This work was supported by the Deutsche Forschungsgemeinschaft (DFG, German Research Foundation) under Germany’s Excellence Strategy within the Cluster of Excellence PhoenixD (EXC 2122, Project ID 390833453). We acknowledge the central computing cluster operated by Leibniz University IT Services (LUIS), which is funded by the DFG (project number INST 187/742-1 FUGG).

\appendix

\section{Numerical calculation of tensors $\hat\alpha_{\rm p}$ and $\hat\alpha_{\rm m}$  \label{sec:appA}}

The elements of the polarizability tensors $\hat\alpha^{\rm p}$ and $\hat\alpha^{\rm m}$ [see Eq.~(\ref{alpha})] are calculated as the following: 
\begin{align}
    &\alpha^{\rm p}_{i}=p_{i}/(\varepsilon_0\varepsilon_{\rm s}E_{0i}),\label{aED}\\
    &\alpha^{\rm m}_{i}=m_{i}/H_{0i}.\label{aMD}
\end{align}
Here, $i=(x,y,z)$, $E_0$ and $H_0$ are the electric and magnetic fields of the incidence wave at the center of the particle (i.e., at the dipole position). Numerical calculation of the components of the electric and magnetic moments $p_{i}$ and $m_{i}$ (e.g., via ANSYS Lumerical's FDTD solver) can be found in the Appendices of Ref.~\cite{allayarov2024multiresonant}.

\begin{figure}[b]
\centering
\includegraphics[width=0.75\linewidth]{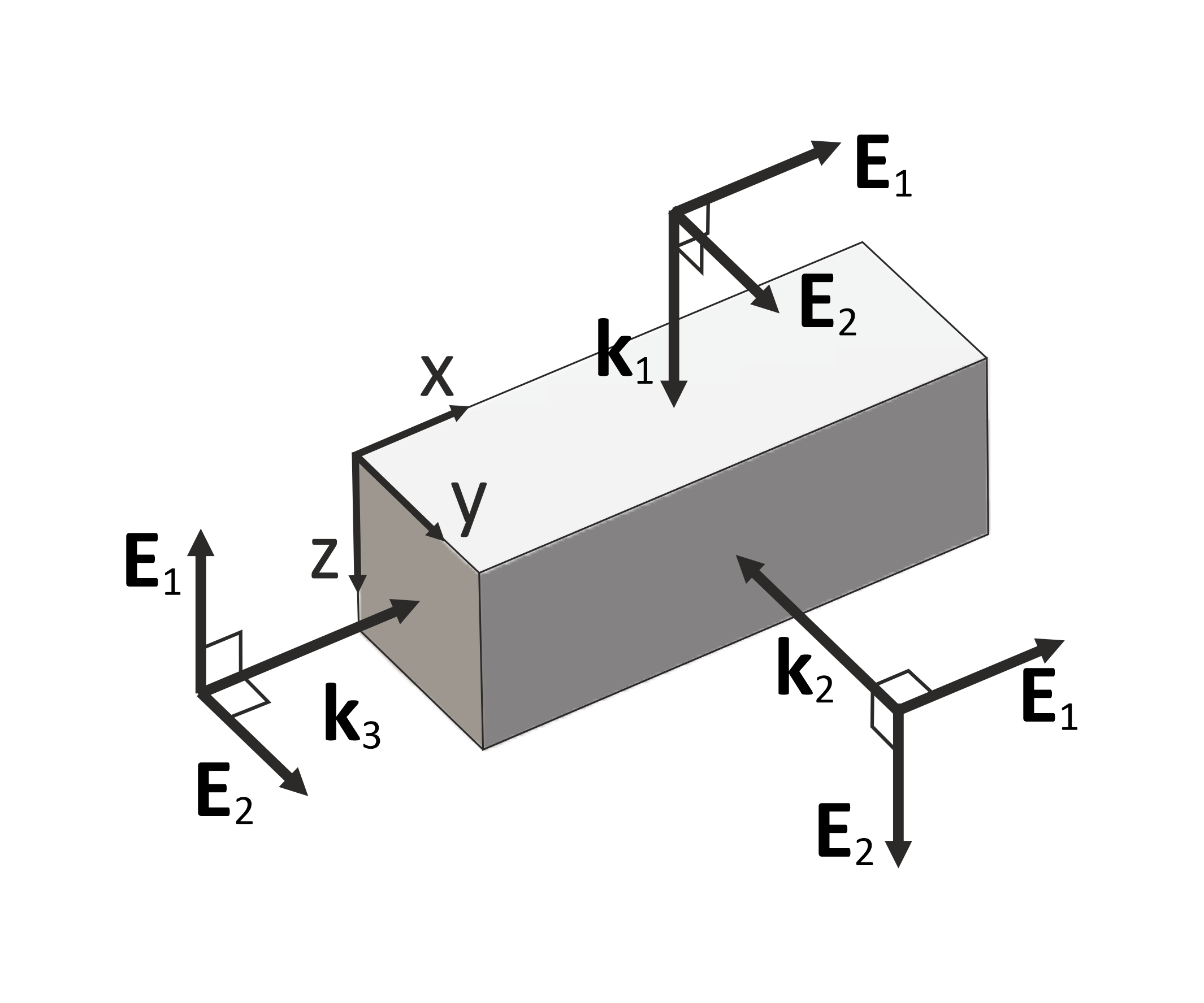}
\caption{Visualization of different excitation conditions of single nanoprism. ${\bf E}$: polarization (e.g., electric field) orientation; ${\bf k}$: excitation direction.}
\label{fig:num_alpha}
\end{figure} 

Since the nanoparticle is non-spherical, polarizability depends on the excitation direction and polarization. In Fig.~\ref{fig:num_alpha}, it is shown three possible excitation directions (${\bf k}_1$, ${\bf k}_2$ and ${\bf k}_3$) of the nanobar and two polarizations (${\bf E}_1$ and ${\bf E}_2$) for each direction. Accordingly, there are six excitation conditions, i.e., two different excitation conditions for each polarizability component:
\begin{align}
    &({\bf k}_1, {\bf E}_1) \quad \text{gives} \quad \alpha^{\rm p}_{x1} \ \text{and} \ \alpha^{\rm m}_{y1},\\
    &({\bf k}_1, {\bf E}_2) \quad \text{gives} \quad \alpha^{\rm p}_{y1} \ \text{and} \ \alpha^{\rm m}_{x1},\\
    &({\bf k}_2, {\bf E}_1) \quad \text{gives} \quad \alpha^{\rm p}_{x2} \ \text{and} \ \alpha^{\rm m}_{z1},\\
    &({\bf k}_2, {\bf E}_2) \quad \text{gives} \quad \alpha^{\rm p}_{z1} \ \text{and} \ \alpha^{\rm m}_{x2},\\
    &({\bf k}_3, {\bf E}_1) \quad \text{gives} \quad \alpha^{\rm p}_{z2} \ \text{and} \ \alpha^{\rm m}_{y2},\\
    &({\bf k}_3, {\bf E}_2) \quad \text{gives} \quad \alpha^{\rm p}_{y2} \ \text{and} \ \alpha^{\rm m}_{z2}.
\end{align}
In our analytical model, we use the tensors with averaged diagonal components:
\begin{align}\label{alphaAVG1}
&\hat\alpha^{\rm d}=\frac{1}{2}\left( {\begin{matrix}
   {\alpha^{\rm d}_{x1}+\alpha^{\rm d}_{x2}} & 0 & {0} \cr
    0& {\alpha^{\rm d}_{y1}+\alpha^{\rm d}_{y2}} & {0} \cr
   {0} & {0} & {\alpha^{\rm d}_{z1}+\alpha^{\rm d}_{z2}} \cr
\end{matrix}
} \right), 
\end{align}
where ${\rm d}= ({\rm p},{\rm m})$.

\bibliography{References}

\end{document}